\documentclass[prl,reprint,groupedaddress]{revtex4-1}
\usepackage{amsmath}
\usepackage{physics}
\usepackage{graphicx}
\usepackage{hyperref}
\usepackage{times}
\usepackage{xcolor}

\newcommand{\subref}[2]{\hyperref[#1]{\ref*{#1}(#2)}}
\definecolor{refblue}{HTML}{2E2E91}
\hypersetup{colorlinks=true, citecolor=refblue, urlcolor=refblue, linkcolor=refblue}

\begin{document}

\title{Engineering Quantum States of Matter for Atomic Clocks in Shallow Optical Lattices}

\author{Ross B. Hutson}
\email{ross.hutson@colorado.edu}
\affiliation{JILA, NIST and University of Colorado, 440 UCB, Boulder, Colorado 80309, USA}
\affiliation{Department of Physics, University of Colorado, 390 UCB, Boulder, CO 80309, USA}
\author{Akihisa Goban}
\affiliation{JILA, NIST and University of Colorado, 440 UCB, Boulder, Colorado 80309, USA}
\affiliation{Department of Physics, University of Colorado, 390 UCB, Boulder, CO 80309, USA}
\author{G. Edward Marti}
\affiliation{JILA, NIST and University of Colorado, 440 UCB, Boulder, Colorado 80309, USA}
\affiliation{Department of Physics, University of Colorado, 390 UCB, Boulder, CO 80309, USA}
\author{Lindsay Sonderhouse}
\affiliation{JILA, NIST and University of Colorado, 440 UCB, Boulder, Colorado 80309, USA}
\affiliation{Department of Physics, University of Colorado, 390 UCB, Boulder, CO 80309, USA}
\author{Christian Sanner}
\affiliation{JILA, NIST and University of Colorado, 440 UCB, Boulder, Colorado 80309, USA}
\affiliation{Department of Physics, University of Colorado, 390 UCB, Boulder, CO 80309, USA}
\author{Jun Ye}
\affiliation{JILA, NIST and University of Colorado, 440 UCB, Boulder, Colorado 80309, USA}
\affiliation{Department of Physics, University of Colorado, 390 UCB, Boulder, CO 80309, USA}
\date{\today}

\begin{abstract}
We investigate the effects of stimulated scattering of optical lattice photons on atomic coherence times in a state-of-the art ${}^{87}\mathrm{Sr}$ optical lattice clock.
    Such scattering processes are found to limit the achievable coherence times to less than $12~\mathrm{s}$ (corresponding to a quality factor of $1 \times 10^{16}$), significantly shorter than the predicted $145(40)~\mathrm{s}$ lifetime of ${}^{87}\mathrm{Sr}$'s excited clock state.
We suggest that shallow, state-independent optical lattices with increased lattice constants can give rise to sufficiently small lattice photon scattering and motional dephasing rates as to enable coherence times on the order of the clock transition's natural lifetime.
Not only should this scheme be compatible with the relatively high atomic density associated with Fermi-degenerate gases in three-dimensional optical lattices, but we anticipate that certain properties of various quantum states of matter can be used to suppress dephasing due to tunneling.
\end{abstract}

\maketitle

Owing to dramatic improvements in both the precision and accuracy of atomic spectroscopy over the last decade~\cite{Ludlow2008, Bloom2014, McGrew2018},
there is growing interest in the use of atomic clocks as quantum sensors in tests of fundamental physics~\cite{Blatt2008, Kolkowitz2016, VanTilburg2015, Pruttivarasin2015, Sanner2018}.
Recent demonstrations of spectroscopic techniques, which are immune to local oscillator noise, promise to dramatically improve the precision of such tests~\cite{Chou2011, Nemitz2016, Schioppo2016, Marti2018}.
In the absence of local oscillator noise, frequency measurements of a single atom follow a binomial distribution and, for Ramsey spectroscopy~\cite{Ramsey1956}, are spread about its true transition frequency $\omega_0$ by an amount $(\omega_0 T)^{-1}$, given in fractional frequency units with $T$ being the coherent evolution time.
In the absence of entanglement, interrogation of a sample of $N$ atoms with an experimental cycle time $T_\mathrm{cyc.}$ results in a quantum projection noise (QPN) limit~\cite{Itano1993},
\begin{equation}
    \sigma_\text{QPN} = \frac{1}{\omega_0 T}\sqrt{\frac{T_\mathrm{cyc.}}{N}}~.
\end{equation}
That is, one wants to increase the interrogation time and use a larger number of atoms in order to reduce the measurement noise.

To date, the lowest reported QPN limit ($\sigma_\mathrm{QPN} = 1.5\times10^{-17} / \sqrt{\mathrm{Hz}}$) was achieved using a  fermi-degenerate gas of $N \approx 10^4$ ${}^{87}\mathrm{Sr}$ ($\omega_0 \approx 2\pi \times 429~\mathrm{THz}$) atoms loaded into the Mott-insulating regime of a three-dimensional (3D) optical lattice~\cite{Campbell2017, Marti2018}.
In these experiments, coherence times were found to be less than $12~\mathrm{s}$ and presumed to be limited by Raman scattering of photons from the deep optical lattice~\cite{Martin2013, Dorscher2018}.
While these scattering processes may be reduced by operating in a shallower optical potential, one then introduces site-to-site tunneling as an additional dephasing mechanism~\cite{Lemonde2005, Kolkowitz2017, Bromley2018}.

In this Letter, we discuss a solution that simultaneously addresses both the lattice photon scattering and tunneling induced dephasing problems in 3D optical lattice clocks: shallow optical lattices with increased lattice constants, $a$.
We find that not only should the decreased kinetic energies in the ground band of such a lattice be sufficient to suppress motional dephasing in a single atom picture, but additionally, for a nuclear-spin polarized Fermi gas at half-filling, inter-electronic-orbital interactions should provide an additional mechanism for reducing motional dephasing rates.
In such a system, atom numbers on the order of $N = 10^5$ and coherent interrogation times up to $T = 140~\mathrm{s}$ seem readily achievable and correspond to a QPN limit of $\sigma_\mathrm{QPN} < 10^{-19}/\sqrt{\mathrm{Hz}}$.

Before describing the details of our proposal, we build upon previous work which investigated trap depth dependent depopulation of the  $5s5p~{}^3P_0$ excited clock state ($\ket{e}$) in ${}^{87}\mathrm{Sr}$ one-dimensional optical lattices as a signature of the Raman scattering problem~\cite{Martin2013, Dorscher2018}. By leveraging the improved control over motional degrees of freedom~\cite{Campbell2017} and imaging techniques~\cite{Marti2018} available in a Fermi-degenerate 3D optical lattice clock, we additionally investigate the corresponding loss of Ramsey fringe contrast.

A spin-polarized degenerate fermi gas is created by evaporatively cooling atoms in an equal mixture of the $m_F = -5/2, \ldots, 9/2$ magnetic sublevels of the $5s^2~{}^1S_0$ electronic ground state ($\ket{g}$) before a focused laser beam, detuned from the $5s5p~{}^3P_1$ intercombination line, provides a state-dependent potential, removing nearly all but the $m_F=9/2$ atoms from the trap.
Approximately $2 \times 10^3$ atoms with a temperature of $20 \%$ of the Fermi temperature are then loaded from the running wave optical dipole trap into a cubic optical lattice.
Each arm ($i = x,y,z$) of the lattice is formed by a retroreflected laser at the magic wavelength ($\lambda_\mathrm{magic}$)~\cite{Ye2008}, and is characterized by a variable depth $V_i$ and a lattice constant $a_0 = \lambda_\mathrm{magic}/2 = 407~\mathrm{nm}$.
The $i=z$ lattice arm is oriented along both gravity and an applied $0.5~\mathrm{mT}$ magnetic bias field.
We perform an additional step of spin purification by coherently driving $\ket{g, m_F=9/2} \rightarrow \ket{e, m_F=9/2}$ with $\lambda_\mathrm{clk} = 698~\mathrm{nm}$ clock light, propagating along the $i=x$ lattice axis, then removing all remaining $\ket{g}$ atoms by cycling on the $5s^2~{}^1S_0 \leftrightarrow 5s6p~{}^1P_1$ transition with resonant $461~\mathrm{nm}$ light.

For the excited state lifetime measurement, we insert a variable hold time before a series of $5~\mu\mathrm{s}$ pulses of $461~\mathrm{nm}$ light form an absorption image of the $\ket{g}$ atoms on a CCD camera, providing a count of the $\ket{g}$ atoms, $N_g$, while also removing the imaged atoms from the trap.
We obtain a count of the remaining atoms, $N_{\tilde{e}}$, by optically pumping $5s5p~{}^3P_0,{}^3P_2 \rightarrow 5s5p~{}^3P_1$ with light resonant on the $5s5p~{}^3P_0,{}^3P_2\leftrightarrow5s6s~{}^3S_1$ transitions at $679~\mathrm{nm}$ and $707~\mathrm{nm}$.
Atoms then rapidly decay to the ground state, via the $21~\mu\mathrm{s}$ lived $5s5p~{}^3P_1$~\cite{Nicholson2015}, where they are subsequently imaged with $461~\mathrm{nm}$ light.
We note that this readout method counts not only atoms in $\ket{e, m_F=9/2}$, but all atoms in the metastable $5s5p~{}^{3}P_0,{}^{3}P_2$ manifold in the quantity $N_{\tilde{e}}$.
The decay of the excited population $\rho_{\tilde{e}\tilde{e}} = N_{\tilde{e}} / (N_g + N_{\tilde{e}})$ is then fit to extract a $1/e$ lifetime.

\begin{figure}
    \includegraphics[width=3.375in]{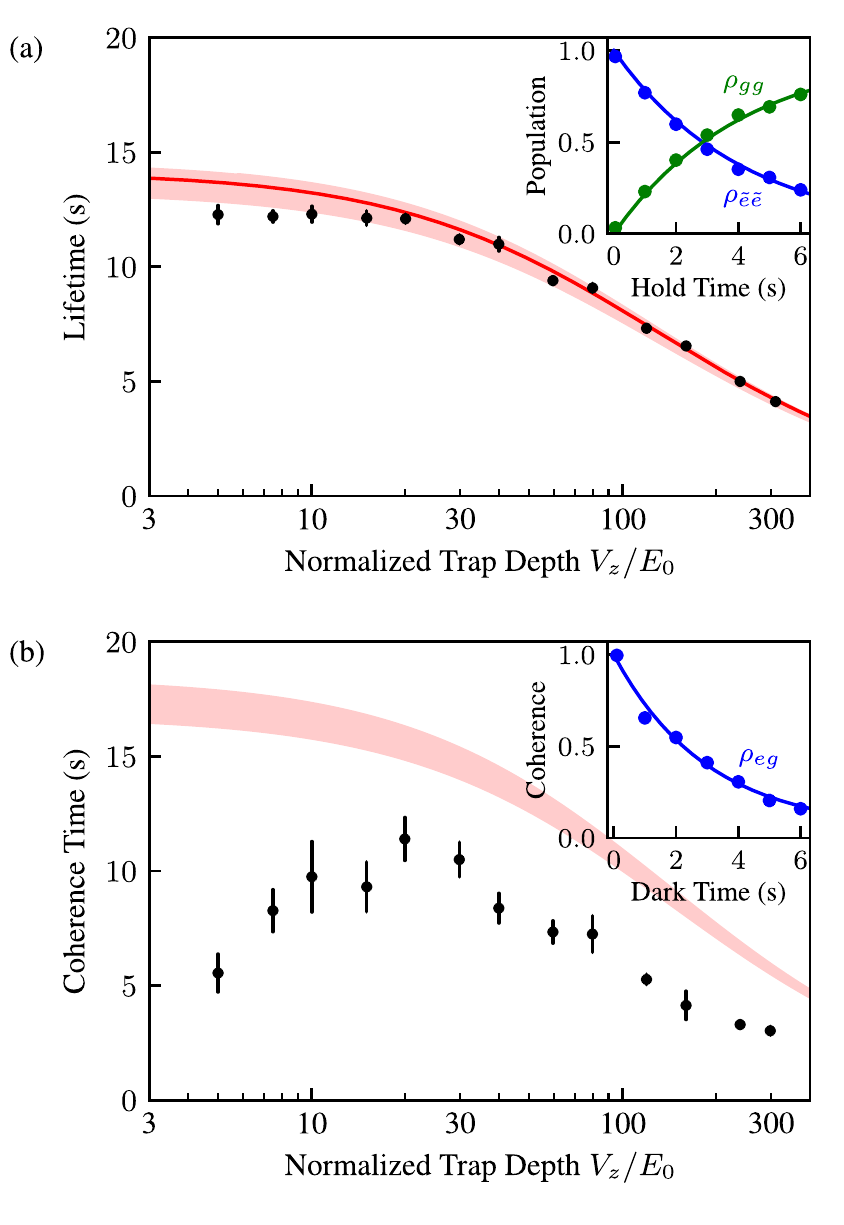}
    \caption{
        Measured (a) lifetimes and (b) coherence times.
        Experimental data are shown in black points along with $1\sigma$ error bars.
        Results of the master equation simulation are shown as shaded red regions.
        The solid red line in (a) represents the fitted decay rate from Ref.~\onlinecite{Dorscher2018}.
        [horizontal lattice depths]
    }
    \label{fig:lifetime-and-contrast}
\end{figure}

These lifetimes are measured for various lattice depths, $V_z$, ranging from $5E_r$ to $310E_r$, while fixing $V_x = 59(2) E_r$ and $V_y = 70(2)E_r$, where $E_r = h^2 / 8 m a_0^2 \approx 3.5~h\cdot\mathrm{kHz}$ is the lattice photon recoil energy, $h$ the Planck constant, and $m$ the atomic mass.
Fig.~\subref{fig:lifetime-and-contrast}{a} shows the trap depth dependence of the extracted lifetimes.
We find the measured lifetimes to be significantly shorter than the predicted $\tau_0 = 145(40)~\mathrm{s}$ natural lifetime~\cite{Boyd2007}, yet largely consistent with numerical simulations with no free parameters (shaded red region) in which two-photon Raman transitions, stimulated by the lattice light, distribute atoms amongst the $5s5p~{}^{3}P_J$ manifold where the atoms can then spontaneously decay to the ground state from $5s5p~{}^{3}P_1$.
The vacuum limited lifetime of atoms prepared in $\ket{g}$ is independently measured to be $> 100~\mathrm{s}$.
An energy level diagram depicting the Raman scattering processes, and the master equation used in the simulation can be found in Ref.~\onlinecite{Supplemental}.

Such scattering events are detrimental to clock operation as they destroy the coherence $\rho_{eg}$ between the two clock states~\cite{Cohen-Tannoudji1992}.
Using imaging spectroscopy~\cite{Campbell2017,Marti2018}, we observe this loss in coherence as a reduction in the Ramsey fringe contrast for increasing dark time, $T$.
The contrast decay at a given lattice depth is then fit to extract a $1/e$ coherence time.
The results of such measurements are shown in Fig.~\subref{fig:lifetime-and-contrast}{b} for the same lattice conditions as in Fig.~\subref{fig:lifetime-and-contrast}{a}.
The observed coherence times are found to scale proportionally to $V_\mathrm{tot.}^{-1}$ ($V_\mathrm{tot.} = \sum_i V_i$) for $V_z > 20E_r$, yet they fall significantly below the predicted decoherence rate due to Raman scattering (shaded red region)~\cite{Supplemental}.

This suggests that other, lattice depth dependent, decoherence mechanisms are present in the system.
Rayleigh scattering is not expected to directly contribute as a dephasing mechanism since the scattering amplitudes are identical for both clock states in a magic wavelength trap~\cite{Uys2010,Martin2013}.
However, both Raman and Rayleigh scattering processes can heat atoms out of the ground band of the lattice~\cite{Gerbier2010} at which point, we suspect, they are able to tunnel around and dephase through contact interactions.

For $V_z < 20 E_r$, coherence times are seen deviate from the $V_\mathrm{tot.}^{-1}$ scaling and instead rapidly decay.
This decay is accompanied by a loss in atom number which we attribute to significant tunneling rates along the $i=z$ lattice and inelastic collisions~\cite{Bishof2011}.
This demonstrates the difficulty in overcoming the Raman scattering problem in conventional optical lattice clocks.
One would like to operate in an optical trap shallow enough to make scattering induced decoherence rates comparable to the natural lifetime --- one requires $V_i \lesssim 4E_r$ for ${}^{87}\mathrm{Sr}$ --- but then one finds additional, tunneling enabled dephasing mechanisms due to the increased kinetic energy scale.

\begin{figure}
    \includegraphics[width=3.375in]{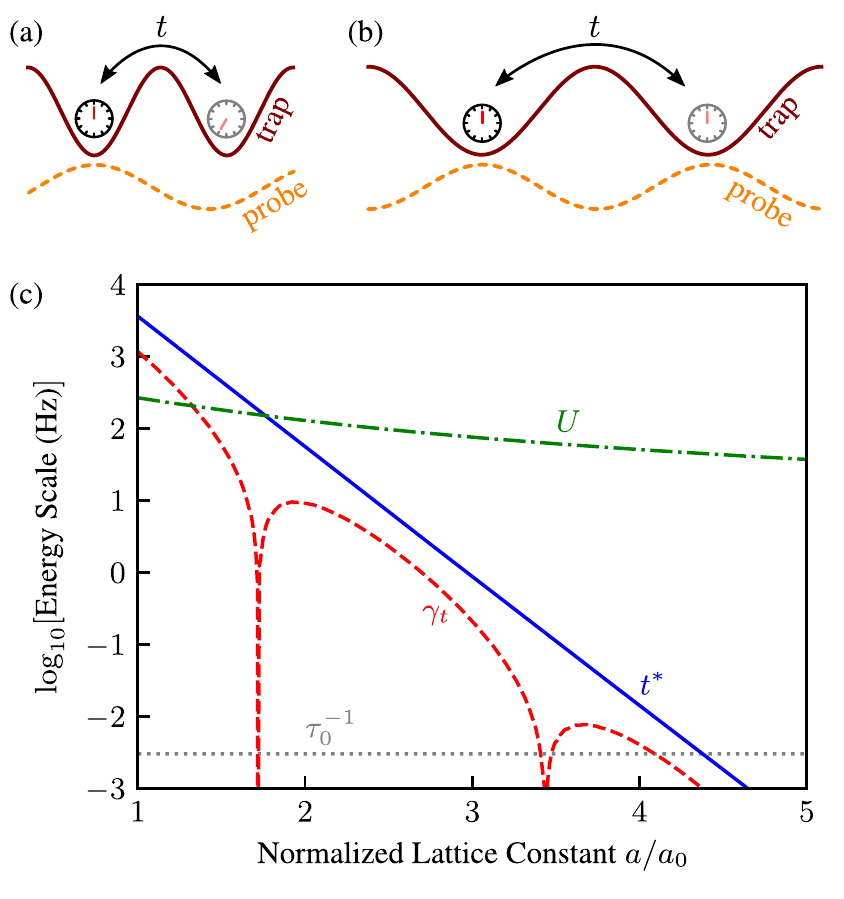}
    \caption{
        (a) Motional dephasing in a conventional, $a/\lambda_\mathrm{clk} \approx 0.6$ lattice.
        An atom in an equal superposition of $\ket{g}$ and $\ket{e}$ is depicted as the face of a clock where the position of the hand reflects the relative phase between the atomic superposition and the local oscillator.
        Upon tunneling to an adjacent lattice site, the relative phase changes by an amount $\phi \approx 1.2 \pi$.
        (b) Motional dephasing can be eliminated by matching the lattice constant to the probe wavelength.
        An atom then sees the same local oscillator phase at each lattice site. 
        (c) Calculated motional dephasing rates ($\gamma_t$), kinetic energies ($t^\ast = 12 t$), and interaction energies ($U$) as a function of lattice spacing, $a$, in a $V = 4E_r$ lattice.
        The horizontal grey line represents the inverse lifetime ($\tau_0^{-1}$) of the strontium clock transition.
        Single particle motional effects are suppressed below the natural decay rate for $a \gtrsim 2~\mu\mathrm{m}$.
    }
    \label{fig:long-wavelength-lattice}
\end{figure}

We address this issue in the following proposal.
Hereafter, we assume a uniform lattice, $V_i = V$, and for the time being, a non-interacting gas.
For a Ramsey type experiment in an inertial reference frame, tunneling at a rate $t$ along the direction of the probe laser results in a loss in contrast of the spectroscopic signal according to $\rho_{eg} = |\mathcal{J}_0(4tT\sin(\phi/2))|$, $\mathcal{J}_0$ being the zeroth order Bessel function of the first kind~\cite{Bromley2018}, and $\phi = 2 \pi a / \lambda_\mathrm{clk}$, the site-to-site phase shift of the clock light where we now allow for a variable lattice constant, $a$, as depicted in Fig.~\subref{fig:long-wavelength-lattice}{a} and Fig.~\subref{fig:long-wavelength-lattice}{b}.
For the purpose of comparing different energy scales of the system, we define the argument of the Bessel function,
\begin{equation}
    \gamma_t = 4 t |\sin(\phi/2)|~,
\end{equation}
as the ``motional dephasing rate''.
As a practical example, such a variable spacing lattice can be engineered, while restricting the wavelength of the trapping light to be magic by interfering the lattice beams at an arbitrary angle, $\theta$, giving a spacing $a = \lambda_\mathrm{magic} / |2 \sin(\theta/2)|$~\cite{Morsch2001, Huckans2009, Al-Assam2010}, or with an optical tweezer array~\cite{Norcia2019}.

Under the harmonic approximation, the tunneling rate for fixed $V$ scales exponentially with $a$ as~\cite{Bloch2008}
\begin{equation}
    \frac{t(a)}{t(a_0)} =  \sqrt{\frac{a_0}{a}}\exp[\sqrt{\frac{V}{E_r}}\left(1 - \frac{a}{a_0}\right)]~.
\end{equation}
One can think of this intuitively as a change in the lattice constant rescaling the lattice recoil energy, $E_r \rightarrow (a_0 / a)^2 E_r $, such that the lattice depth in units of the new recoil energy can be made quite large for modest increases in $a/a_0$.
Numerical values of $\gamma_t$ and the total kinetic energy , $t^\ast = 12t$, for a $V = 4E_r$ lattice are shown in Fig.~\subref{fig:long-wavelength-lattice}{c}.
For a sufficiently large lattice constant, the atomic limit ($t \rightarrow 0$) is achieved where tunneling related effects can be neglected. We find that both $\gamma_t$ and $t^\ast$ are suppressed below $\tau_0^{-1}$ for lattice spacings $a \gtrsim 2~\mu\mathrm{m}$.

Additionally, $\gamma_t$ is found to sharply drop to zero upon matching the condition $a / \lambda_\mathrm{clk} \mod 1 = 0$.
These resonances can be understood in a momentum space picture where the clock photon recoil is matched to a reciprocal lattice vector and thus absorption or emission of a clock photon leaves each atom's motional state unchanged.
In this case, for a nuclear-spin polarized gas at half filling, the system behaves as a band insulator throughout clock spectroscopy as the indistinguishability of all atoms is preserved.
This scheme, however, requires an accuracy in $\theta$ beyond the $2 \times 10^{-5}$, and $2 \times 10^{-2}$ levels for the $a/\lambda_\mathrm{clk} = 1, 2$ configurations, respectively.
Throughout this range of parameters ($1 < a / a_0 < 5$) the lattice bandgap is greater than $h\cdot2~\mathrm{kHz}$, and the effective Rabi coupling is suppressed by no more than 60\% of the bare Rabi coupling such that carrier resolved spectroscopy is easily achievable; line-pulling effects from off-resonant excitation of motional sidebands can be suppressed below the $10^{-19}$ level for Rabi frequencies below 10 Hz, as shown in Fig.~S4~\cite{Supplemental}.

Many-body effects arise through an on-site interaction energy $U$ parameterized by the anti-symmetric inter-electronic-orbital $s$-wave scattering length, $a_{eg^-} = 69.1 (0.9)~a_B$~\cite{Goban2018}, where $a_B$ is the Bohr radius.
This energy scale decreases algebraically with an increasing lattice constant,
\begin{equation}
    \frac{U(a)}{U(a_0)} = \left(\frac{a_0}{a}\right)^{3/2}~,
\end{equation}
such that for sufficiently large lattice spacings, the system enters the Mott-insulating regime ($t^\ast / U \ll 1$)~\cite{Jordens2008,Schneider2008}.
Here, for a sufficiently cold gas at half-filling, the only available excitations below the energy gap $U$ are of the order of the superexchange energy $J = 4t^2 / U$.
Thus we expect, for a sufficiently weak probe pulse, the motional dephasing rates to be suppressed by a factor of $t / U$ as compared to the non-interacting case. 
Numerical values for $U$ in a $V = 4 E_r$ lattice are shown in Fig.~\subref{fig:long-wavelength-lattice}{c}.

\begin{figure}
    \includegraphics[width=3.375in]{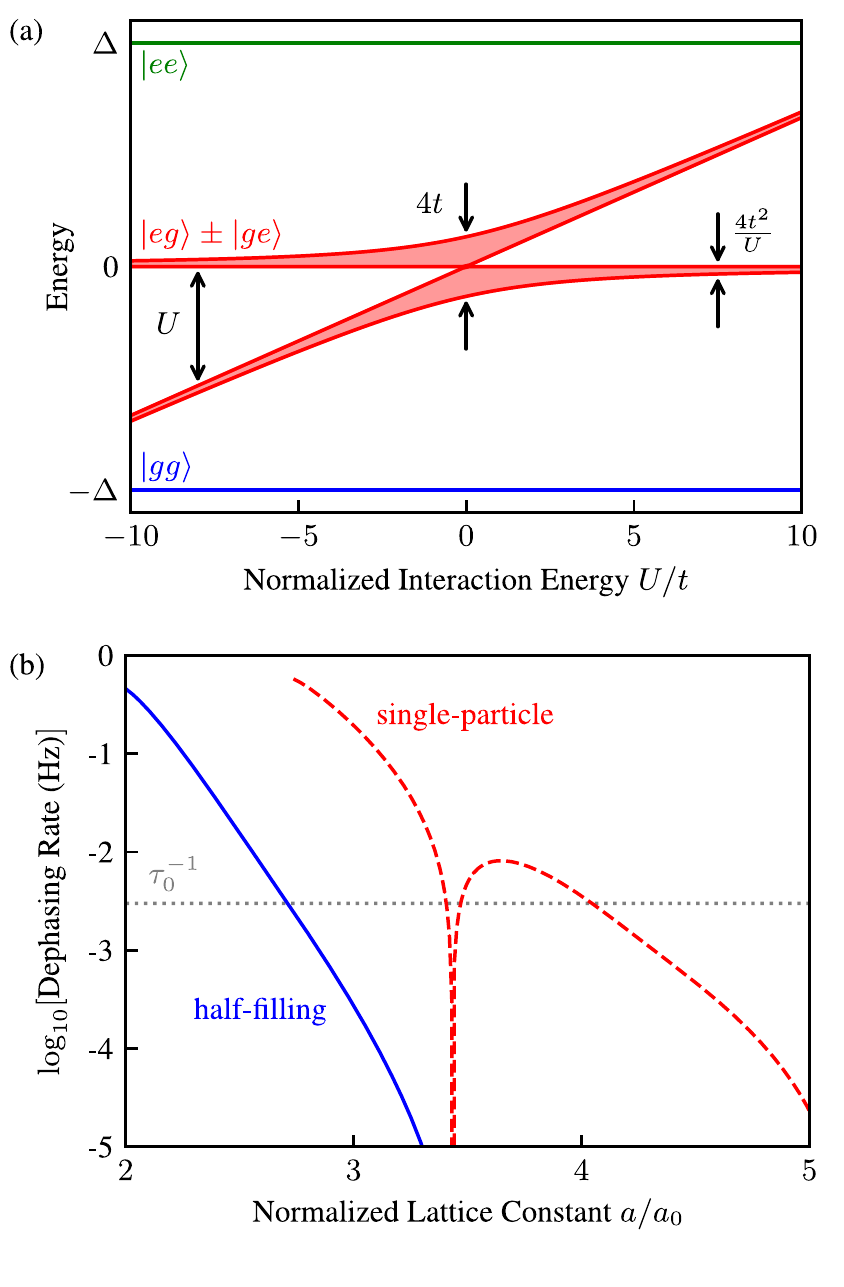}
    \caption{
    (a) Energy spectrum of Eqn.~\ref{eqn:double-well-hamiltonian} at half-filling.
    States with zero, one, and two atoms in $\ket{e}$ are shown as blue, red, and green lines, respectively.
    The $\ket{gg}$ and $\ket{ee}$ states are non-interacting due to the Pauli exclusion principle.
    The $\ket{eg} \pm \ket{ge}$ states are spread by twice the Bloch band width at $U=0$. 
    Whereas in the Mott-insulating regime ($U/t \gg 1$), an energy gap $U$ opens up and a pair of weakly interacting states become spectroscopically resolvable.
    (b) Dephasing rates, as given by Eqn.~\ref{eqn:broadening}, for one (red dashed line) and two (solid blue line) atoms in a double well potential.
    The tunneling rates and interaction strengths as a function of lattice spacing are taken from Fig.~\subref{fig:long-wavelength-lattice}{c} and inserted into the double well Hamiltonian with $\Omega / 2\pi = 0.5~\mathrm{Hz}$.
    The curves are not plotted for $4t > \Omega$ where the analogy between the double well system and an infinite lattice breaks down as the discrete levels of the finite sized system become resolved.
    }
    \label{fig:double-well}
\end{figure}

We investigate these effects with a ``toy model'' consisting of a double well potential.
We assign the following Hamiltonian in the rotating wave approximation,
\begin{equation}
\begin{aligned}
    H/\hbar = &\sum_x
    \Big[
        \frac{\Delta}{4} \left(n_{x,e} - n_{x,g}\right)
        + \frac{U}{2}n_{x,g}n_{x,e}\\
    &   + \frac{\Omega}{2} e^{i\phi_x}c_{x,e}^\dag c_{x,g}
    \Big]
    - t \sum_\sigma c_{\mathrm{L}, \sigma}^\dag c_{\mathrm{R}, \sigma} + \mathrm{H.c.}
\end{aligned}
\label{eqn:double-well-hamiltonian}
\end{equation}
to such a system.
Here $c_{x, \sigma}^\dag$ ($c_{x, \sigma}$) creates (destroys) a fermion with internal state $\sigma \in \{g, e\}$ in well $x \in \{\mathrm{L}, \mathrm{R}\}$, $n_{x,\sigma} = c^\dag_{x,\sigma}c_{x,\sigma}$, $\Delta = \omega - \omega_0$ is the difference between the frequency of the driving field $\omega$ from the atomic resonance, $\omega_0$, $\Omega$ is the Rabi coupling strength, and $\phi_x = 2 \pi \frac{a}{\lambda_\mathrm{clk}}\delta_{x, \mathrm{R}}$ is the site-dependent phase shift of the clock light, with $\delta_{i,j}$ being the Kronecker delta function.
The two atom spectrum of this Hamiltonian with $\Omega = 0$, is shown in Fig.~\subref{fig:double-well}{a}.

We simulate Ramsey spectroscopy of one and two atoms in the double well by numerically integrating the Schr\"odinger equation.
A resonant ($\Delta = 0$) $\pi/2$-pulse with $\Omega \gg t$ places each atom in an equal superposition of ground and excited electronic states.
For $a / \lambda_\mathrm{clk} \mod 1 \neq 0$, this pulse also changes the system's motional state.
During field-free evolution ($\Omega = 0$), the different motional states beat against  each other causing a dephasing of the spectroscopic feature.
We quantify this effect with the following relation,
\begin{equation}
    \gamma = \sqrt{\langle H^2\rangle - \langle H\rangle^2}.
    \label{eqn:broadening}
\end{equation}

For a single atom, this quantity approximates the dephasing rate in an infinite lattice, $\gamma_t$, falling off with an envelope proportional to $t$ as shown in Fig.~\subref{fig:double-well}{b} (red dashed line).
For two atoms, we observe that as one begins to resolve the interaction energy ($\Omega \ll U$) the dephasing rate becomes proportional to the superexchange energy, falling off with an envelope proportional to $t^2 / U$ as shown in Fig.~\subref{fig:double-well}{b} (solid blue line).
While the exact numerical prefactor differs slightly from what one would get for an infinite lattice~\cite{Essler2005}, the general conclusion is the same: the minimum lattice spacing such that $\gamma \tau_0 < 1$ is significantly relaxed as compared to the non-interacting case.

We have identified scattering of lattice photons as a dominant decoherence mechanism in a state-of-the-art 3D optical lattice clock and proposed a number of ways in which quantum materials may be engineered to overcome such limits.
The improved clock stability associated with longer coherence times will directly enable new searches for time variation of fundamental constants and tests of general relativity on sub-mm length scales.
Additionally, the shallow optical potentials described in this Letter will help reduce systematic clock shifts related to the traps themselves, especially terms that are nonlinear in trap depth~\cite{Porsev2018, Ushijima2018}.
Furthermore, increasing the distance between atoms should aid in reducing systematic clock shifts arising from collective radiative effects~\cite{Chang2004,Bromley2016}.
Future work can investigate the use of atomic collisions to create metrologically useful entanglement~\cite{He2019}.

\begin{acknowledgements}
We acknowledge stimulating conversations with S. Kolkowitz and S. L. Campbell, and also thank C. J. Kennedy, M. A. Perlin, A. M. Rey for careful reading of the manuscript. This work is supported by NIST, DARPA, AFOSR-MURI, and Grant No. NSF-1734006. A.G. is supported by a postdoctoral fellowship from the Japan Society for the Promotion of Science.
\end{acknowledgements}

\bibliography{references}

\end{document}


\renewcommand{\theequation}{S\arabic{equation}}
\renewcommand{\thefigure}{S\arabic{figure}}

\title{Supplemental Material: Engineering Quantum States of Matter for Atomic Clocks in Shallow Optical Lattices}

\author{Ross B. Hutson}
\affiliation{JILA, NIST and University of Colorado, 440 UCB, Boulder, Colorado 80309, USA}
\affiliation{Department of Physics, University of Colorado, 390 UCB, Boulder, CO 80309, USA}
\author{Akihisa Goban}
\affiliation{JILA, NIST and University of Colorado, 440 UCB, Boulder, Colorado 80309, USA}
\affiliation{Department of Physics, University of Colorado, 390 UCB, Boulder, CO 80309, USA}
\author{G. Edward Marti}
\affiliation{JILA, NIST and University of Colorado, 440 UCB, Boulder, Colorado 80309, USA}
\affiliation{Department of Physics, University of Colorado, 390 UCB, Boulder, CO 80309, USA}
\author{Lindsay Sonderhouse}
\affiliation{JILA, NIST and University of Colorado, 440 UCB, Boulder, Colorado 80309, USA}
\affiliation{Department of Physics, University of Colorado, 390 UCB, Boulder, CO 80309, USA}
\author{Christian Sanner}
\affiliation{JILA, NIST and University of Colorado, 440 UCB, Boulder, Colorado 80309, USA}
\affiliation{Department of Physics, University of Colorado, 390 UCB, Boulder, CO 80309, USA}
\author{Jun Ye}
\affiliation{JILA, NIST and University of Colorado, 440 UCB, Boulder, Colorado 80309, USA}
\affiliation{Department of Physics, University of Colorado, 390 UCB, Boulder, CO 80309, USA}
\date{\today}

\maketitle

\section{I. Master Equation}
\subsection{Derivation}
Here we extend the results of Ref. 26 to include auxillary atomic levels.
We consider a Hilbert space, $\mathcal{H} = \mathcal{H}_\mathrm{S} \otimes \mathcal{H}_\mathrm{R}$, comprised of the electronic states of ${}^{87}\mathrm{Sr}$, $\psi_S \in \mathcal{H}_\mathrm{S}$, and photonic states $\psi_R \in \mathcal{H}_\mathrm{R}$ spanned by the polarizations $\bm{\epsilon}$, and wavevectors $\mathbf{k}$ of a quantization volume $\mathcal{V}$.
The electronic Hamiltonian is given by
\begin{equation}
    H_S = \hbar \sum_{i \in \mathcal{H}_\mathrm{S}} \omega_i \ket{i}\bra{i}
\end{equation}
\begin{figure}
    \includegraphics[width=3.375in]{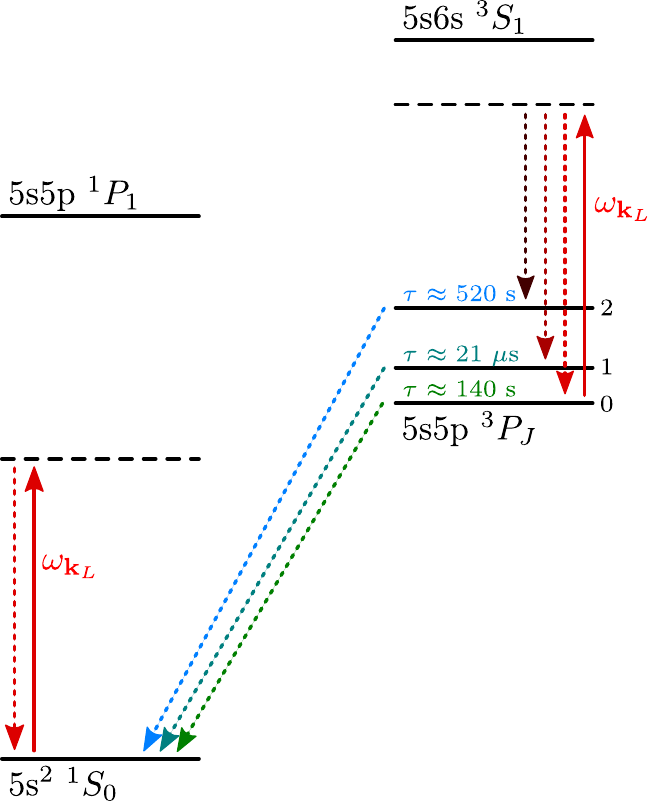}
    \caption{
        A few relevent energy levels of ${}^{87}\mathrm{Sr}$.
    }
    \label{fig:energy-levels}
\end{figure}
where $\omega_i$ are the atomic Bohr frequencies, a few of which are shown schematically in Fig.~\ref{fig:energy-levels}.
The photonic Hamiltonian is
\begin{equation}
    H_R = \hbar \sum_{\mathbf{k},\bm{\epsilon}} \omega_\mathbf{k} \left[n_{\mathbf{k}, \bm{\epsilon}}  + \frac{1}{2}\right]
\end{equation}
where $n_{\mathbf{k}, \bm{\epsilon}} = a_{\mathbf{k}, \bm{\epsilon}}^\dag a_{\mathbf{k}, \bm{\epsilon}}$ is the photon number operator, $a^\dag_{\mathbf{k},\bm{\epsilon}}$ ($a_{\mathbf{k},\bm{\epsilon}}$) creates (destroys) a photon with wavevector $\mathbf{k}$ and polarization $\bm{\epsilon}$, and $\omega_\mathbf{k} = c |\mathbf{k}|$ with $c$ the speed of light.
Under the dipole approximation, these degrees of freedom are then coupled according to
\begin{equation}
    \label{eqn:dipole}
    H_{SR} = \mathbf{d}\cdot\mathbf{\mathcal{E}}~.
\end{equation}
Here $\mathbf{d}$ is the electric dipole operator acting on the electronic degree of freedom, and
\begin{equation}
    \mathbf{\mathcal{E}} = i\sum_{\mathbf{k},\bm{\epsilon}} 
    \sqrt{\frac{\hbar\omega_\mathbf{k}}{2\epsilon_0 \mathcal{V}}} \left[a_{\mathbf{k}\bm{\epsilon}}e^{-i \omega_\mathbf{k} t + i\mathbf{k}\cdot\mathbf{r}}
     - a_{\mathbf{k}\bm{\epsilon}}^\dag e^{i \omega_\mathbf{k} t - i\mathbf{k}\cdot\mathbf{r}} \right]\bm{\epsilon}~.
\end{equation}
We consider a photonic state, $\ket{L}$ where one mode $(\mathbf{k}_L, \bm{\epsilon}_L)$ is in an off-resonant coherent state corresponding to a classical electric field of strength $\mathcal{E}_L$, while all other modes are thermally populated ($n_{\mathbf{k},\bm{\epsilon}} \ll 1$).
To first order, Eqn.~\ref{eqn:dipole} only couples electronic states of opposite parity, however higher order effects can couple levels of equal parity. 
We capture this behavior with an effective Hamiltonian, $H'_{SR}$, which we derive by integrating the Schr\"odinger equation to second order.
The atom-light interaction then has two effects.
First, the energies of the electronic states, $\ket{i}$, are shifted according to 
\begin{equation}
    \bra{i;L}H'_{SR}\ket{i;L} = -\frac{1}{4}\alpha^{ii}_{\bm{\epsilon}_L} \mathcal{E}_L^2
\end{equation}
where 
\begin{equation}
    \alpha^{ij}_{\bm{\epsilon}} = \frac{1}{\hbar}\sum_{l}
    \left[\frac{\bra{j}\mathbf{d}\cdot\bm{\epsilon}\ket{l}\bra{l}\mathbf{d}\cdot
    \bm{\epsilon}_L\ket{i}}{\omega_l-\omega_i - \omega_{\mathbf{k}_L}} + \frac{\bra{j}\mathbf{d}\cdot\bm{\epsilon}_L
    \ket{l}\bra{l}\mathbf{d}\cdot\bm{\epsilon}\ket{i}}{\omega_l - \omega_j + \omega_{\mathbf{k}_L}}\right] ~.
\end{equation}
Secondly, other modes of the photon reservoir are populated according to
\begin{equation}
    \bra{j;L}a_{\mathbf{k},\bm{\epsilon}}H'_{SR}\ket{i;L} = -\frac{\mathcal{E}_L}{2}
    \sqrt{\frac{\hbar \omega_\mathbf{k}}{2 \epsilon_0 \mathcal{V}}} \alpha^{ij}_{\bm{\epsilon}}\delta_{\omega_i + \omega_{\mathbf{k}_L},\omega_j + \omega_{\mathbf{k}}}
\end{equation}

This second effect then leads to decoherence in the reduced density matrix $\rho_\mathrm{S} = \Tr_\mathrm{R} \ket{\psi}\bra{\psi}$ according to the master equation,
\begin{align}
    \frac{\partial \tilde{\rho}_S}{\partial t} &\approx - \frac{1}{\hbar^2}\int_0^t \mathrm{d}t'~\Tr_\mathrm{R}\left[H'_\mathrm{SR}(t), \left[H'_\mathrm{SR}(t'), \rho_\mathrm{S}(t')\rho_R\right]\right]\\
    &\approx \frac{1}{2}\sum_{i \neq j} \Gamma_{ij}^\mathrm{in.}\Big[
        2\tilde{\rho}_{ii}\ket{j}\bra{j} -
        \sum_l \big(\tilde{\rho}_{il} \ket{i}\bra{l} + \tilde{\rho}_{li} \ket{l}\bra{i}\big)\Big] - \frac{1}{2}\sum_{i \neq j} \Gamma^\mathrm{el.}_{ij}\tilde{\rho}_{ij}\ket{i}\bra{j}
    \label{eqn:master}
\end{align}
Where the coefficients $\Gamma^x_{ij}$ are given by
\begin{equation}
    \Gamma_{ij}^\mathrm{in.} = \frac{I(\omega_{\mathbf{k}_L} - \omega_{ji})^3}{(4\pi\epsilon_0)^2c^4\hbar^3}\frac{8\pi}{3}\sum_{\bm{\epsilon}}\left|\alpha^{ij}_{\bm{\epsilon}}\right|^2~,
\end{equation}
and
\begin{equation}
    \Gamma_{ij}^\mathrm{el.} = \frac{I\omega_{\mathbf{k}_L}^3}{(4\pi\epsilon_0)^2c^4\hbar^3}\frac{8\pi}{3}\sum_{\bm{\epsilon}}\left|\alpha^{ii}_{\bm{\epsilon}} - \alpha^{jj}_{\bm{\epsilon}}\right|^2~.
\end{equation}
with $I=\epsilon_0 c \mathcal{E}_L^2/2$.
We calculate numerical values of these coefficients for all 100 magnetic sublevels of the metastable manifold $\{5s^2~{}^1S_0, 5s5p~{}^3P_J\}$ of ${}^{87}\mathrm{Sr}$ using the excited state lifetimes found in Refs.~17, 22, and 43-46.

\subsection{Effective Rates}
Handling the full master equation with all 100 magnetic sublevels, can be somewhat cumbersome. Here, we quote effective rates for clock state population decay and decoherence. 
For excited state population decay according to $\partial_t \rho_{ee} = -\Gamma_{ee}\rho_{ee}$, we find 
\begin{equation}
    \Gamma_{ee} = \sum_i\Gamma'_{ee} V_i\left(1 - \sqrt{E_r/4V_i}\right) + \Gamma^0_{ee}
\end{equation}
with $\Gamma'_{ee} = 5.7(3) \times 10^{-4}~\mathrm{s}^{-1}E_r^{-1}$ and $\Gamma^0_{ee} = 9(1) \times 10^{-3}~\mathrm{s}^{-1}$, where $\Gamma^0_{ee}$ takes into account both spontaneous emission and thermal photon depumping via $5s4d~{}^3D_1$~[17].
For clock state decoherence according to $\partial_t \rho_{eg} = -\Gamma_{eg}\rho_{eg}$, we find 
\begin{equation}
    \Gamma_{eg} = \sum_i\Gamma'_{eg} V_i\left(1 - \sqrt{E_r/4V_i}\right) + \Gamma^0_{eg}
\end{equation}
with $\Gamma'_{eg} = 4.0(2) \times 10^{-4}~\mathrm{s}^{-1}E_r^{-1}$, and $\Gamma^0_{eg} = 1.2(1) \times 10^{-2}~\mathrm{s}^{-1}$.
In the above expressions, the non-linear scaling of the scattering rates with lattice depth accounts for the spatial overlap of the atomic density with the laser intensity in the harmonic limit ($V_i/E_r \gg 1)$.

\subsection{Comparison with Blue-detuned Lattice}
It has been proposed~[17] that operation in a blue-detuned magic wavelength lattice, e.g. $\lambda_b \approx 390~\mathrm{nm}$~[47], would be a possible method to reduce Raman scattering induced dephasing due to the atoms being trapped at the lattice intensity minima.
For clock state decoherence according to $\partial_t \rho_{eg} = -\Gamma^b_{eg}\rho_{eg}$, we find 
\begin{equation}
    \Gamma^b_{eg} = \sum_i\Gamma^{b}{}'_{eg} V_i\sqrt{E_b/4V_i} + \Gamma^0_{eg}
\end{equation}
with $\Gamma^b{}'_{eg} = 110(10) \times 10^{-4}~\mathrm{s}^{-1}E_b^{-1}$, and $E_b = h^2 / 2m\lambda_b^2$.
We can then compare the expected dephasing rates in blue versus red detuned lattices by picking $V_i$ to achieve similar tunneling rates along all lattice axes.
For $V_\mathrm{tot.} = 150 E_r$ in the red-detuned case, one requires $V_\mathrm{tot.} \approx 185E_b)$ in the blue-detuned case.
This gives $\Gamma_{eg} \approx 14~\mathrm{s}$ and $\Gamma^b_{eg} \approx 13~\mathrm{s}$ and we thus conclude that simply operating in a blue-detuned lattice does not provide for a significant advantage over the red-detuned case, and will not enable atomic coherence times on the order of $\tau_0$.

\section{II. Line pulling}
\subsection{External Hamiltonian}
\begin{figure}
    \includegraphics[width=3.375in]{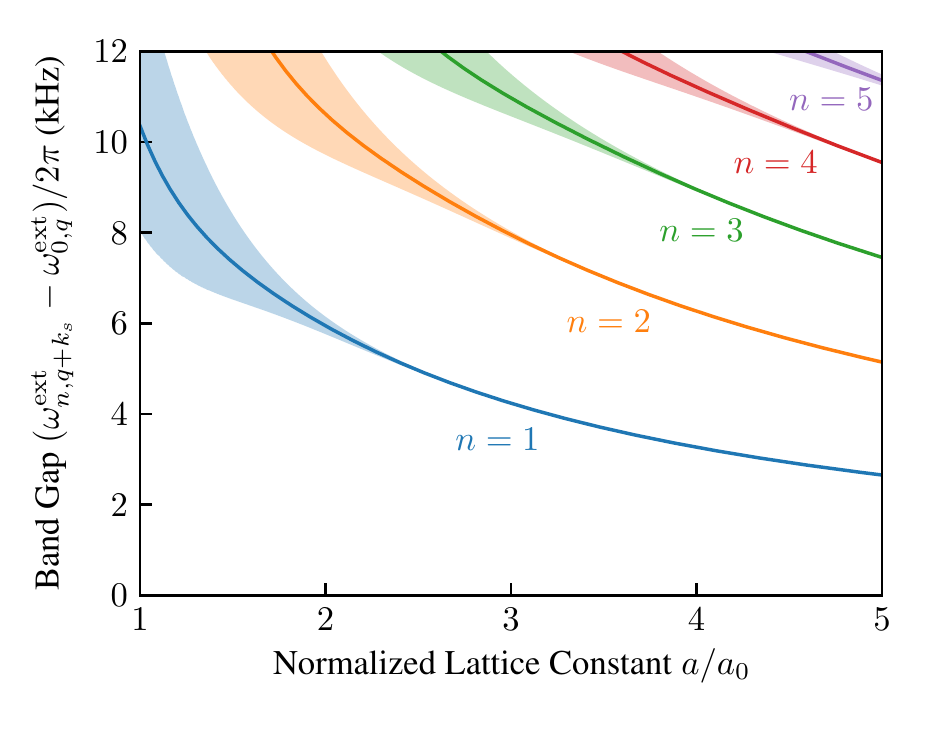}
    \caption{
        Band gaps as a function of lattice spacing in a $V = 4E_r$ lattice.
        Shaded regions show the full range of values for all initial quasi-momenta, $q$.
        Solid lines show mean values.
    }
    \label{fig:band-gaps}
\end{figure}

Here we extend some of the results of Ref.~[18] by allowing for a variable spacing lattice.
The potential due to the $i^\mathrm{th}$ axis of the separable optical lattice can be expressed as
\begin{equation}
    V_i(x_i) = V \sin^2\left(k_l x_i\right)~,
\end{equation}
with $k_l = \pi / a$.
The corresponding Hamiltonian can be given in the plane wave basis $\ket{k}$ according to
\begin{equation}
    \bra{k'}H^\mathrm{ext} \ket{k} = \left(\frac{\hbar^2 k^2}{2m} + \frac{V}{2}\right)\delta_{k',k} - \frac{V}{4}\left(\delta_{k',~k+2k_l} + \delta_{k',~k-2k_l}\right)~.
\end{equation}
Numeric diagonalization yields the eigenstates $\ket{n, q}$, given in terms of the plane waves as
\begin{equation}
    \ket{n, q} = \sum_m C_{n,~q + 2 m k_l} \ket{q + 2 m k_l}~,
\end{equation}
with corresponding energies
\begin{equation}
    \bra{n,q}H^\mathrm{ext}\ket{n, q} = \hbar\omega^\mathrm{ext}_{n, q}~.
\end{equation}
Numerical values of this spectrum are given in Fig.~\ref{fig:band-gaps}.

\subsection{Rabi Coupling}

\begin{figure}
    \includegraphics[width=3.375in]{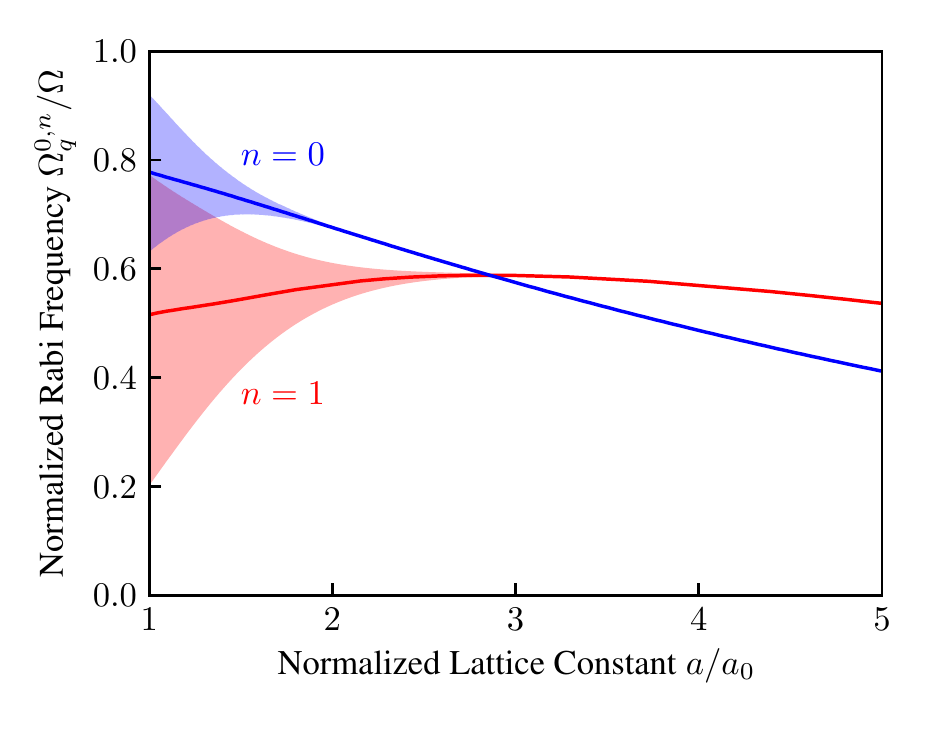}
    \caption{
        Rabi couplings as a function of lattice spacing in a $V = 4E_r$ lattice.
        Shaded regions show the full range of values for all initial quasi-momenta, $q$.
        Solid lines show mean values.
    }
    \label{fig:rabi-frequencies}
\end{figure}

In addition to coupling the internal states, the clock laser couples different motional states according to
\begin{equation}
    \bra{e;n',q'} \mathbf{d}\cdot\mathcal{E}\ket{g;n,q} = \frac{1}{2}\bra{e}\mathbf{d}\cdot\mathcal{E}_0e^{-i \omega t}\ket{g}\bra{n', q'} e^{i k_s x}\ket{n, q} = \frac{\hbar\Omega}{2} \braket{n', q'}{n, q+k_s}
\end{equation}
where $\Omega = \bra{e}\mathbf{d}\cdot\mathcal{E}_0\ket{g}/\hbar$ and $k_s = 2\pi/\lambda_\mathrm{clk}$.
Numerical values for the quantities
\begin{equation}\begin{aligned}
    \Omega^{n,n'}_q &\equiv \frac{1}{\hbar}\sum_{q'}\bra{e;n',q'} \mathbf{d}\cdot\mathcal{E}\ket{g;n,q} = \Omega \sum_{m} C_{n',~q + k_s + 2 m k_l}C_{n, q + 2 m k_l}
\end{aligned}\end{equation}
are given in Fig.~\ref{fig:rabi-frequencies}.

\subsection{Clock Shift}

\begin{figure}
    \includegraphics[width=3.375in]{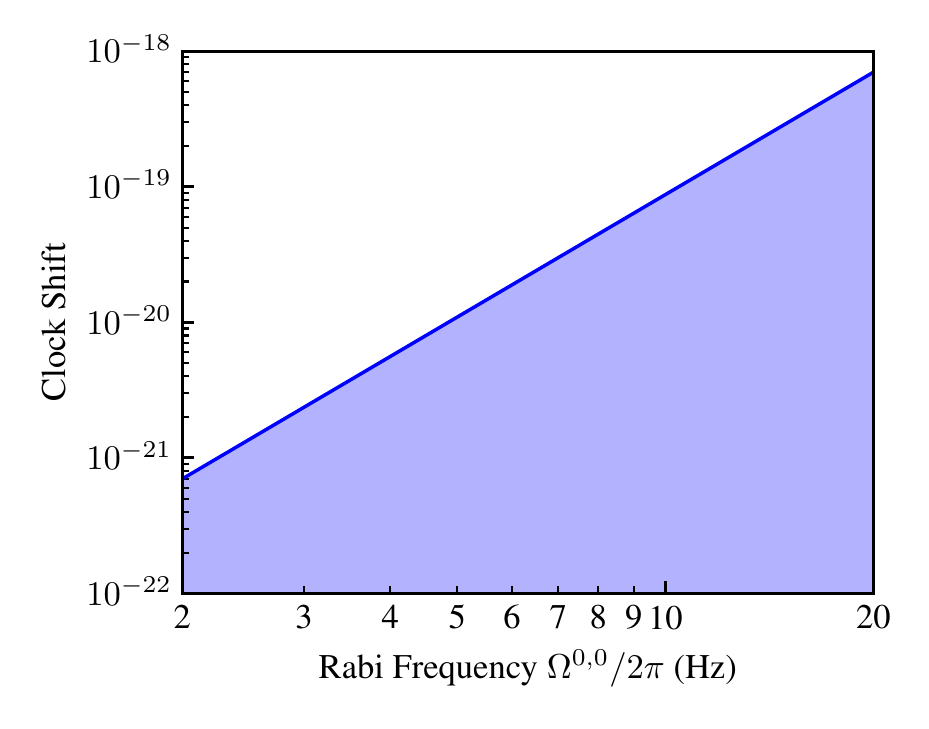}
    \caption{
        Bounds on line pulling clock shifts for the range of parameters considered in the main text.
        Upper bounds generated with $a/a_0=5$, $\omega^\mathrm{ext}_1 - \omega^\mathrm{ext}_0=2.6~\mathrm{kHz}$ and $\Omega^{0,1}/\Omega^{0,0} = 1.3$.
    }
    \label{fig:line-pulling}
\end{figure}

In a Rabi interrogation sequence, an error signal for referencing the clock laser to the atomic transition is typically generated by starting in the state $\ket{\psi_\pm(0)} = \ket{g;n=0}$, applying a pulse of duration $T_\pi = \pi/\Omega^{0,0}$ while dithering the detuning between two values $\Delta_\pm \approx \Delta \pm 0.8 \Omega$ on succesive experimental cycles, then computing the difference in excitation probabilities, $\epsilon(\Delta) = P_+ - P_-$.
The presence of asymmetric sidebands in the atomic response can then result in a non-zero clock shift, $\Delta/\omega_0$ for $\Delta$ satisfying $\epsilon(\Delta) = 0$.

We may calculate the quantities $P_\pm$ by first solving the system of coupled differential equations
\begin{equation}\begin{aligned}
    \partial_t \braket{e;n}{\psi_\pm(t)} &= -\frac{i}{2}\sum_{n'}\Omega^{n,n'}e^{-i(\Delta_\pm + \omega^\mathrm{ext}_{n'} - \omega^\mathrm{ext}_{n})t}\braket{g;n'}{\psi_\pm(t)}\\
    \partial_t \braket{g;n}{\psi_\pm(t)} &= -\frac{i}{2}\sum_{n'}\Omega^{n',n}e^{i(\Delta_\pm + \omega^\mathrm{ext}_{n} - \omega^\mathrm{ext}_{n'})t}\braket{e;n'}{\psi_\pm(t)}~,
\end{aligned}\end{equation}
then summing
\begin{equation}
    P_\pm = \sum_n\left|\braket{e;n}{\psi_\pm(T_\pi)}\right|^2~.
\end{equation}
Here, all dependence on $q$ has been suppressed under the assumption that any variance is negligibly small for $a/a_0 \gtrsim 3$.
Fig.~\ref{fig:line-pulling} gives the range of clock shifts relevant to our proposal, showing that line pulling effects are suppressed below the $10^{-19}$ level for $\Omega / 2 \pi < 10~\mathrm{Hz}$.